\documentclass[twocolumn,aps,pra,floatfix,longbibliography]{revtex4-2}

\usepackage{amsmath}
\usepackage{txfonts}
\usepackage{microtype}
\usepackage{graphicx}
\usepackage{color}
\usepackage{ulem}
\usepackage{bm}
\newcommand{\comm}[1]{}
\usepackage{calrsfs}
\DeclareMathAlphabet{\pazocal}{OMS}{zplm}{m}{n}

\usepackage{mathalfa}

\begin{document}

\title{Role of quantum dynamics in coherent and incoherent radiation during tunneling ionization}

\author{Michael Klaiber}\email{klaiber@mpi-hd.mpg.de}
\author{Karen Z. Hatsagortsyan}\email{k.hatsagortsyan@mpi-hd.mpg.de}
\author{Christoph H. Keitel}
\affiliation{Max-Planck-Institut f\"ur Kernphysik, Saupfercheckweg 1, 69117 Heidelberg, Germany}

\date{\today}

\begin{abstract}

Radiation during  strong-field tunneling ionization is investigated. The spontaneous  as well as the coherent components of the radiation are calculated  describing the ionization dynamics via the strong field approximation and the  role of  the quantum dynamics at tunneling is analyzed. The competition  between  different mechanisms  in different spectral regions is examined. Transition-like radiation (Brunel radiation) is dominated at low-frequencies, Thomson scattering  at the laser frequency, and radiative recombination via the three-step process  at  high-order harmonics. To distinguish the role of the quantum dynamics, simple man Drude models are developed  for the coherent as well as for spontaneous radiation, which are based on the electron trajectory out of the tunneling barrier. The quantum dynamics is shown to modify the coherent Brunel radiation for near-zero-frequencies in asymmetric laser pulses. The significant role of free-free transitions is demonstrated for the spontaneous radiation in the low-frequency region.

\end{abstract}

\maketitle

\section{Introduction}

It is well known that tunneling ionization is accompanied by  high-order harmonic generation (HHG) via the three-step model \cite{Corkum_1993}, when the recombination of the ionized electron wave packet to the bound state at the recollision after the excursion in the laser field and gaining energy results in the emission of a high energy photon \cite{Ferray_1988,Lewenstein_1994,Agostini_2004}. At the recollision, radiation may emerge also due to  electron scattering, rather than recombination, yielding  free-free HHG \cite{Kohler_2010}.
In the macroscopic picture, the ionization of a gas target results in the creation of an electron current  after the interaction with a laser pulse, which can be a source of  low frequency THz radiation, with the emitted frequency  determined by the interaction time
 \cite{Sprangle_2004,Kim_2008,Kim_2009,Dai_2009,Wen_2009,Babushkin_2010a,Babushkin_2010b,
 Babushkin_2011,You_2012,Clerici_2013,Balciunas_2013,Oh_2014,Vvedenskii_2014,Tulsky_2018}.
  The same mechanism can work also in the case of a single atom producing THz radiation in asymmetric laser pulses, e.g. in a two-color laser field  \cite{Babushkin_2017,Babushkin_2022}. The origin of this  radiation is similar to the transition radiation, when the electron velocity \cite{Bolotovskii_1986} (or dielectric constant of a medium where the electron propagates \cite{Ginzburg_1979})  abruptly changes. A similar effect exists in plasma physics -- Brunel radiation \cite{Brunel_1987,Brunel_1990}, when an electron in the plasma   is dragged out of the plasma. In the case of tunneling ionization, the electron is initially in a stationary  bound state, while abruptly starting to move after tunneling.

In the tunneling regime of strong-field ionization, the bound electron tunnels through the barrier created by the laser and the Coulomb field of the atomic core. However, this common simple dynamics can   continue to evolve in a nontrivial way. The electron traversing the classically forbidden region may reflect from the potential barrier surface, propagate back and scatter off the core, and then finally tunnel out of the barrier. Although this nontrivial path has no large probability, it has consequences, inducing the tunneling time delay when interfering with the direct ionization path \cite{Klaiber_2018,Klaiber_2022R}, and it can be responsible for driving Freeman resonances in the strong field regime \cite{Klaiber_2020}, which has recently been confirmed experimentally \cite{Khurelbaatar_2025}. The question arises whether the signatures of the 
quantum dynamics at tunneling, in particular, those of the under-the-barrier recollisions, can be observed in the radiation accompanying tunneling ionization.

In contrast to the coherent radiation discussed above,   emission of spontaneous low-order harmonic radiation during the bound-continuum transition with the origin similar to Brunel  radiation has been recently shown in~\cite{Milosevic_2023}, with the typical frequency determined by the Keldysh tunneling time \cite{Keldysh_1965}. Finally, harmonic radiation at tunneling ionization arises also from  Thomson scattering of the laser wave off the ionized free electron wave packet, with higher harmonics than the fundamental one  suppressed in the nonrelativistic regime.

Coherent radiation should be distinguished from incoherent radiation of the same origin. For instance, the spontaneous recombination of a high-energy electron to an atomic bound state results in a photon emission of a random phase, which is an incoherent process \cite{Sundaram_1990}. However, when the recombining electron is released from the bound state and recombines into the same bound state driven by a laser field, the HHG  is coherent with the phase determined by the coherent superposition of the ground state wave function with the ionization continuum wave packet. Similarly, the coherent part of radiation due to Compton scattering of a strong laser wave from an ionized electron wave packet \cite{Krekora_2002,Chowdhury_2005} is phase-matched in forward direction, while the incoherent radiation has a broad angular distribution \cite{Peatross_2008,Corson_2011,Tarbox_2015,Ware_2016,Pan_2019}. For the incoherent process, in contrast to the coherent one, the final state of the electron does not exist initially. The coherent radiation intensity scales as $\sim {\cal N}^2$ with the number of emitters~${\cal N}$, due to the phase matching of the radiation, while for the incoherent radiation interference is suppressed and the total intensity scaling is $\sim {\cal N}$ \cite{Marcuse_1971}. The coherent radiation can be calculated using the quantum mechanical expectation value of the current density within the classical radiation theory \cite{Peatross_2008}.

In this paper we investigate the role of the under-the-barrier dynamics for the spontaneous and coherent Brunel radiation during tunneling ionization. The ionization quantum dynamics is described by strong field approximation (SFA) which implicitly includes the under-the-barrier path. Additionally, we develop  simple man Drude models for the coherent as well as for spontaneous radiation, which include only the electron trajectory out of the tunneling barrier. Comparison of the results of the Drude model with those of SFA allows us to single out the contribution of the under-the-barrier path in the radiation spectra. We analyze  how the different mechanisms of the radiation during the strong field ionization are scaled in different spectral regions of emission and the feasibility of their observation in a coherent or incoherent form.

The harmonics emitted from a gas target is the result of a multi-atom contribution, and therefore, are dominated by the coherent radiation. For the observation of the incoherent radiation either a single atom experiment should be carried out (when each time a single atom existed in the interaction region), or the emission should be measured at a stray direction far from the phase-matching one \cite{Ware_2016}.

The structure of the paper is the following. In Sec.~\ref{sec:coh} the coherent radiation during tunneling ionization is calculated using the SFA. Spontaneous emission at the tunneling ionization is discussed in Sec.~\ref{sec:spont}. Relation of our calculations to an experiment is considered in Sec.~\ref{sec:experiment}. We give an order of magnitude estimation of radiation intensities in Sec.~\ref{sec:estim}. Our conclusions are given in Sec.~\ref{sec:concl}.

\section{Coherent radiation during tunneling ionization}\label{sec:coh}

\subsection{Radiation via SFA current}

We calculate coherent radiation emitted during tunneling ionization of a laser-driven hydrogenlike atom in the nonrelativistic regime. The dynamics of the tunneling ionization is treated by the SFA, which provides the current density of the ionizing electron. The latter is employed in Maxwell equations to calculate the intensity of coherent radiation.

The strong-field ionization dynamics is described by the Schr\"odinger equation
\begin{eqnarray}
i\partial_t|\Psi (t)\rangle=\hat{H} \,|\Psi (t)\rangle,
\end{eqnarray}
with the Hamiltonian
\begin{eqnarray}
\hat{H}=\hat{\mathbf{p}}^2/2+\hat{V}(r)+\mathbf{r}\cdot\mathbf{E}(t)
\end{eqnarray}
where $\hat{V}(\mathbf{r})=2\pi/\kappa\delta(\mathbf{r})(1+\partial_r)$ is the  regularized atomic zero-range potential, and $\mathbf{E}(t)$ the laser electric field. The interaction with the laser field is described in the length gauge which is the most suitable physically in the SFA \cite{Milosevic_2006,Scully_Zubairy_1997}. Atomic units are used throughout.

We seek the solution for the electron wave function $\Psi$ via the iterative SFA expansion  \cite{Becker_2002}:
\begin{eqnarray}
|\Psi(t)\rangle=|\psi^{(0)}(t)\rangle+|\psi^{(1)}(t)\rangle+|\psi^{(2)}(t)\rangle,
\end{eqnarray}
where the zeroth-order term $|\psi^{(0)}\rangle$ coincides with the unperturbed bound  state $|\psi^{(a)}\rangle$ in the atomic potential. The first-order SFA wave function describes the direct tunneling path:
\begin{eqnarray}
|\psi^{(1)}(t)\rangle=-i\int^t dt'\int d^3\mathbf{p}\,|\psi^{(f)}_\mathbf{p}(t)\rangle\langle\psi^{(f)}_\mathbf{p}(t')|\mathbf{r}\cdot \mathbf{E}(t')|\psi^{(a)}(t')\rangle\nonumber\\
\end{eqnarray}
with the continuum Volkov state $|\psi^{(f)}(t)\rangle$ in the laser field \cite{Volkov_1935},
and in the second-order SFA \cite{Becker_2002}:
\begin{eqnarray}
|\psi^{(2)}(t)\rangle &=&-\int^t dt''\int^{t'} dt'\int d^3\mathbf{p}\int d^3\mathbf{q}\, |\psi^{(f)}_\mathbf{p}(t)\rangle\nonumber\\&&\times\langle\psi^{(f)}_\mathbf{p}(t'')|V|\psi^{(f)}_\mathbf{q}(t'')\rangle\langle\psi^{(f)}_\mathbf{q}(t')|\mathbf{r}\cdot \mathbf{E}(t')|\psi^{(a)}(t')\rangle.
\end{eqnarray}
The states $|\psi^{(1)}\rangle$ and $|\psi^{(2)}\rangle$ describe the  electron dynamics in the continuum and at recollisions, including the under-the-barrier dynamics, and the motion in the laser field after leaving the barrier.

From previous studies it is known \cite{Peatross_2008} that the coherent radiation can be calculated
interpreting the quantum current density as a classical one within the classical radiation
theory.
We calculate the quantum current density for the tunneling electron  $\mathbf{j}(\mathbf{r},t)=\frac{1}{2}\{\Psi(\mathbf{r},t)^*\hat{\mathbf{p}}\Psi(\mathbf{r},t)+\rm c.c.\}$,  interpret it as a classical current and use it to calculate the emitted \textit{coherent} radiation \cite{Marcuse_1971} within the standard gauge invariant representation of Lienard-Wiechert radiation  theory. From the latter the photon emission probability reads \cite{Landau_2}:
\begin{eqnarray}
dw=\frac{\alpha}{(2\pi)^2}\frac{d^3\mathbf{k}}{k^3}\left|\int dt \int \,d^3\mathbf{r} \,\mathbf{k}\times \mathbf{j}(\mathbf{r},t)e^{i\omega t- i\mathbf{k}\cdot\mathbf{r}}\right|^2,
\end{eqnarray}
where $\alpha$ is the fine structure constant, $\omega$ and $\mathbf{k}$ are the emitted photon energy and momentum, respectively, and  $k=|\mathbf{k}|$. In the nonrelativistic regime the dipole approximation for the emission is relevant, $\mathbf{k}\cdot\mathbf{r}\ll 1$. Then, the angular integration of the emitted photon probability can be carried out yielding:
\begin{eqnarray}
\frac{dw}{d\omega}=\frac{2\alpha\omega}{3\pi c^2}\left|\int dt \,\mathbf{J}(t) \exp(i\omega t)\right|^2,
\label{dw}
\end{eqnarray}
where $\mathbf{J}(t)=\int d^3\mathbf{r} \,\mathbf{j}(\mathbf{r},t)=\langle\Psi (t)\left|\hat{\mathbf{p}}\right|\Psi (t)\rangle$ is the expectation value of the electron probability current, and $c$ is the speed of light.
We can identify different probability currents $\mathbf{J}(t)=\mathbf{J}_{01}(t)+\mathbf{J}_{10}(t)+\mathbf{J}_{11}(t)$,
where
\begin{eqnarray}
\mathbf{J}_{01}(t)&=&\langle\psi^{(0)}(t)\left|\hat{\mathbf{p}}\right|\psi^{(1)}(t)\rangle,\nonumber\\
 \mathbf{J}_{10}(t)&=&\langle\psi^{(1)}(t)\left|\hat{\mathbf{p}}\right|\psi^{(0)}(t)\rangle,\\
 \mathbf{J}_{11}(t)&=&\langle\psi^{(1)}(t)\left|\hat{\mathbf{p}}\right|\psi^{(1)}(t)\rangle.\nonumber
\end{eqnarray}
 The radiation via the combination of $\mathbf{J}_{01}$ and $\mathbf{J}_{10}$ emerges due to the recombination to the bound state and, therefore, describes three-step HHG, including below-threshold harmonics.  The radiation via $\mathbf{J}_{11}$ term includes the effect of the electron transition to the continuum and the continuum dynamics, which will be the main point of our study. The second-order terms $\mathbf{J}_{02}$, $\mathbf{J}_{20}$ describe the contribution to the radiation due to the under-the-barrier recollisions, which are usually small in the net radiation, and the observation of its signatures requires more subtle analysis, see Sec.~\ref{sec:spont}C.

The expression for the photon emission probability of Eq.~(\ref{dw}) can be equivalently derived via the quantum
perturbation theory, treating the interaction of the electron with the photon field in the velocity gauge $H_{\rm rad}=\hat{\mathbf{A}} (t)\cdot\hat{\mathbf{p}}$ by perturbation:
\begin{eqnarray}
dw&=&\left|m_{\mathbf{k}}\right|^2d^3\mathbf{k} \\
m_{\mathbf{k}}&=&-i \int dt\langle\Psi(t),1_\mathbf{k}|\hat{\mathbf{A}} (t)\cdot\hat{\mathbf{p}}|\Psi(t),0_\mathbf{k}\rangle,\nonumber
\end{eqnarray}
where $\hat{\mathbf{A}}  =-\hat{\mathbf{E}} (t)/\omega$ is the vector potential operator for the emitted photon, and $ \hat{\mathbf{E}}(\mathbf{r},t)=\sum_\mathbf{k} \sqrt{\omega}/(2\pi) \hat{\mathbf{e}} \left(a_\mathbf{k}e^{i \mathbf{k}\cdot\mathbf{r}-i\omega t}+a^+_\mathbf{k}e^{-i \mathbf{k}\cdot\mathbf{r}+i\omega t}\right)$ is the  field of the emitted photons in second quantization, with the creation and annihilation operators $a^+_\mathbf{k}$ and $a_\mathbf{k}$. As the classical radiation theory is gauge invariant, we take its result as a benchmark to choose the gauge of the quantum
perturbation theory with respect to the radiation Hamiltonian. Thus, we use different gauges in description of different interactions: the length gauge to describe the electron interaction with the laser field, justified by the SFA experience,
and the velocity gauge for the perturbative treatment of the interaction with the emitted photon field, justified by the matching with the Lienard-Wiechert theory.

\subsection{Simple man Drude model for Brunel radiation}

The radiation via the current $\mathbf{J}_{11}$ in Eq.~(\ref{dw}) describes radiation during tunneling ionization due to free-free transitions. This includes, first of all, Brunel radiation, which arises due to the bound electron abruptly starting to move after tunneling, and secondly, Thomson scattering of the laser wave on the ionized free electron, which contributes mostly at the laser frequency. The Bremsstrahlung in the applied short-range atomic potential is suppressed. We are mostly interested in Brunel radiation. We will see that the current $\mathbf{J}_{11}$ contains  contributions  from the continuum dynamics after the tunneling, as well as from the under-the-barrier path. To clearly single out the effect of the under-the-barrier  dynamics for the Brunel radiation,
we introduce a simple man model  dubbed as Drude model in Ref.~\cite{Babushkin_2017}, and simulate the radiation only due to the continuum dynamics excluding the under-the-barrier contribution.

In this model we evaluate the ionized electron probability current in the continuum $\mathbf{J}_{\rm D}(t)$ at the time instant $t$, consisting of the tunnel ionized electron probability contribution at previous times $t'<t$, multiplied by the laser driven velocity after the tunneling. We use the ionization probability via the Perelomov-Popov-Terent'ev (PPT) theory \cite{Perelomov_1967a,Popov_2004u}, including nonadiabatic corrections:
\begin{eqnarray}
  W(t')=\frac{\kappa^2}{2} \frac{|\mathbf{E}(t')|}{E_a}\exp\left[-\frac{2E_a\left(1-\frac{\gamma^2}{10}\right)}{3|\mathbf{E}(t')|}\right],
\end{eqnarray}
where $E_a=\kappa^3$ is the atomic field, $I_p$ the atomic ionization energy, $\kappa=\sqrt{2I_p}$, 
$\gamma=\kappa\omega_0/E_0$ Keldysh parameter.  The laser driven velocity after the tunneling is 
\begin{equation}
  \mathbf{v}(t,t')=\mathbf{A}(t)-\mathbf{A}(t').
\end{equation}
Thus, the  ionized electron probability current in the continuum reads:
\begin{eqnarray}
\label{Bab}
\mathbf{J}_{\rm D}(t)= \int^t dt'\, \mathbf{v}(t,t')W(t').
\end{eqnarray}
Using $\mathbf{J}_{\rm D}(t)$, the radiation spectra with the Drude model is calculated using classical radiation theory via Eq.~(\ref{dw}).

Within this model, the  radiation has a spike in the near-zero-frequency  domain in the case of an asymmetric pulse, which has a simple explanation. Initially, the electron is at rest at the core, then it tunnels out  during the current formation time $\tau$, and is accelerated in the continuum due to the laser field up to its final velocity $v$. The current associated with this motion can be parameterized as
\begin{eqnarray}
\label{erf}
J(t)\sim v \,\,[1+\text{erf}(t/\tau)],
\end{eqnarray}
with the Fourier component of the current $J_\omega \sim v \exp(-\omega^2\tau^2/4)/\omega$. Thus, from Eq.~(\ref{dw})
\begin{eqnarray}
\frac{dw}{d\omega} \sim \frac{v^2}{\omega}e^{- \omega^2\tau^2/2} ,
\end{eqnarray}
and the spectrum has  a singularity at zero-frequency. Note that the Drude model exaggerates the singularity because it does not account accurately for the formation time $\tau$ of the current, because the error-function of Eq.~(\ref{erf}) is replaced by a Heaviside-theta function.

\begin{figure*}
  \begin{center}
\includegraphics[width=0.45\textwidth]{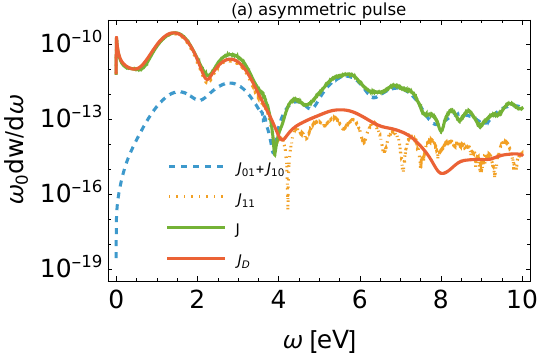}
\includegraphics[width=0.45\textwidth]{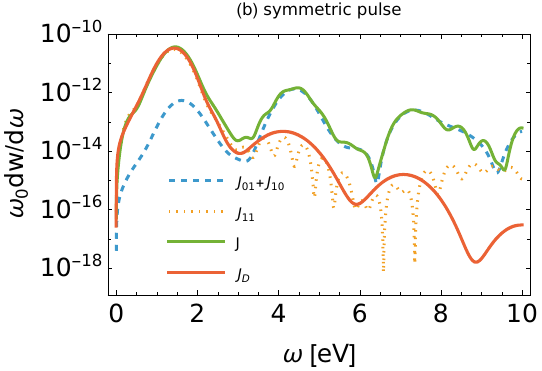}
   \caption{Coherent radiation spectra during tunneling ionization with $E_0=0.1$, $\omega_0=0.05$, and $N=12$: (a) in the asymmetric laser pulse ${\cal A}=0.1$ [Eq.~(\ref{alpha})], (b) in the  symmetric laser pulse ${\cal A}=0$. Comparison of the Brunel radiation via $J_{11}$ with the three-step HHG ($J_{01}+J_{10}$)  and with the Drude model ($J_{\rm D}$). The total radiation with the full current $J=J_{01}+J_{10}+J_{11}$ is given for comparison.   }
\label{JJ}
\end{center}
\end{figure*}
\begin{figure*}
  \begin{center}
\includegraphics[width=0.45\textwidth]{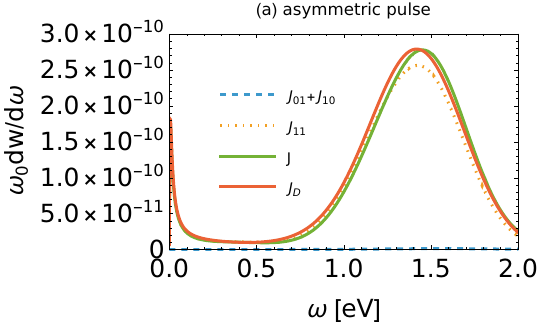}
\includegraphics[width=0.45\textwidth]{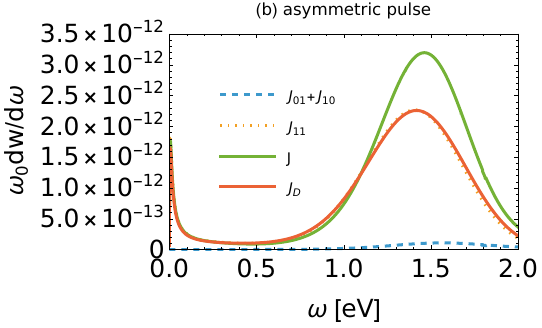}
\includegraphics[width=0.45\textwidth]{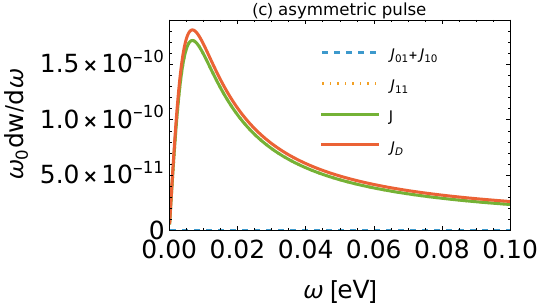}
\includegraphics[width=0.45\textwidth]{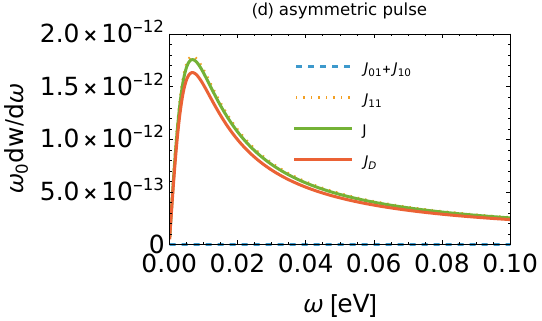}
   \caption{Coherent radiation spectra during tunneling ionization in the asymmetric laser pulse: (first column) $E_0 = 0.1 $ [zoomed-in view Fig.~\ref{JJ}(a)]; (second column) $E_0 = 0.075$. Comparison of the Brunel radiation via $J_{11}$ with the three-step HHG ($J_{01}+J_{10}$)  and  with the Drude model ($J_{\rm D}$). }
\label{JJ2}
\end{center}
\end{figure*}

\subsection{Coherent Brunel radiation}

\subsubsection{Calculation of the current}

We have carried out the calculation of the coherent radiation spectra in the case of a symmetric and an asymmetric  few-cycle laser pulse defined by their displacement
\begin{eqnarray}
\label{alpha}
{\boldsymbol \alpha}(t) = -\hat{\mathbf{e}}_x\frac{E_0}{\omega_0^2} \left[\cos(\omega_0 t) +  \mathcal{A} \sin(2 \omega_0 t)\right] \cos(\omega_0 t/N)^4 ,
\end{eqnarray}
where $\alpha''(t)=A'(t)=-E(t)$, $E_0$ and $\omega_0$ are the laser field amplitude and frequency, respectively, $N=12$,  ${\cal A}$ is the asymmetry parameter of the laser pulse, ${\cal A}=0$ for the symmetric monochromatic pulse, and ${\cal A}=0.1$ for the asymmetric two-color one. In the calculation of currents, the time integrations are performed numerically, while all other integrals are evaluated analytically  in the applied 3D zero-range potential case.

The current component $\mathbf{J}_{01}(t)$ reads
\begin{eqnarray}
\mathbf{J}_{01}(t)=-i\int^t dt'\int d^3\mathbf{p}\,\langle\psi^{(0)}(t)\left|\hat{\mathbf{p}}\right|\psi^{(f)}_\mathbf{p}(t)\rangle\langle\psi^{(f)}_\mathbf{p}(t')|\mathbf{r}\cdot \mathbf{E}(t')|\psi^{(a)}(t')\rangle,\nonumber
\end{eqnarray}
which after the saddle-point integration over the momentum is expressed as follows:
\begin{eqnarray}
\label{J01t}
\mathbf{J}_{01}(t)=-i\int^t dt' \left(\frac{2\pi i}{t-t'-i\epsilon  }\right)^{3/2}\mathbf{m}_p^*(\mathbf{p}_s,t)m_x(\mathbf{p}_s,t')
\end{eqnarray}
where $\mathbf{p}_s=-[\boldsymbol{\alpha}(t)-\boldsymbol{\alpha}(t')]/(t-t')$ is the saddle-point  momentum, and the singularity is handled using $\epsilon =0.4/\kappa^2$  \cite{Suarez_2015}. Here, the matrix elements are defined
\begin{eqnarray}
\mathbf{m}_p(\mathbf{p},t)&=&\int d^3\mathbf{r} \psi^f_\mathbf{p}(\mathbf{r},t)^* \hat{\mathbf{p} }\psi^a (\mathbf{r},t)=\frac{ \sqrt{\kappa } (\mathbf{p}+\mathbf{A}(t))}{\pi  \left[\kappa ^2+(\mathbf{p}+\mathbf{A}(t))^2\right]^2}\nonumber\\
&&\times\exp[ip^2/2t+i\mathbf{p}\cdot\boldsymbol{\alpha}(t)+i\beta(t)+i\kappa^2/2t],\nonumber\\
m_x(\mathbf{p},t)&=&\int d^3\mathbf{r} \psi^f_\mathbf{p}(\mathbf{r},t)^* \hat{\mathbf{r}}\cdot\mathbf{E}(t)\psi^a (\mathbf{r},t)\nonumber\\
&=&-2i \,\mathbf{m}_p(\mathbf{p},t)\cdot \mathbf{E}(t),
\end{eqnarray}
with $\beta=\int dt \,A^2(t)/2$.

The current component $\mathbf{J}_{11}(t)$ reads
\begin{eqnarray}
 \mathbf{J}_{11}(t)&=&-\int^t dt'\int^t dt''\int d^3\mathbf{p}\int d^3\mathbf{p}'\langle\psi^{(a)}(t')|\mathbf{r}\cdot \mathbf{E}(t')|\psi^{(f)}_\mathbf{p}(t')\rangle\nonumber\\
 &\times &
 \langle\psi^{(f)}_\mathbf{p}(t)\left|\hat{\mathbf{p}}\right|\psi^{(f)}_{\mathbf{p}'}(t)\rangle\langle\psi^{(f)}_\mathbf{p}(t'')|\mathbf{r}\cdot \mathbf{E}(t'')|\psi^{(a)}(t'')\rangle,\nonumber \\
\label{J11t}
 &=&\int d^3 \mathbf{p}(\mathbf{p}+\mathbf{A}(t))\left|\int^t dt'm_x(\mathbf{p},t')\right|^2.
\end{eqnarray}
  The $t$-integration in Eq.~(\ref{J01t}), $t$- and $p$-integrations in Eq.~(\ref{J11t}), and the Fourier transform in Eq.~(\ref{dw}) are calculated numerically.

Further, because of a finite duration of the laser pulse with sharp time boundaries, unphysical contributions arise \cite{Klaiber_2022edge} in the temporal Fourier transformation of the currents, which are removed by the following procedure. In the interval $(t_i,t_f)$ the time-integration for the Fourier transform is carried out numerically, while beyond it analytically:
\begin{eqnarray}
\int_{-\infty}^\infty dt f(t) =\frac{f(t_i)^2}{f'(t_i)}+\int ^{t_f}_{t_i}dt f(t)-\frac{f(t_f)^2}{f'(t_f)},
\end{eqnarray}
where $t_i$ and $t_f$ are the  switching-on and -off times of the laser pulse,  respectively, using
\begin{eqnarray}
\int^{t_i}_{-\infty}dt f(t) 
&\approx&\int^{t_i}_{-\infty}dt \exp[\ln(f(t_i)+f'(t_i)(t-t_i))]=\frac{f(t_i)^2}{f'(t_i)},\nonumber
\end{eqnarray}
with the function $f$ being the integrand of the Fourier-transformation  $f(t)=J(t)\exp[i\omega t-\epsilon t]$, where the parameter $\epsilon=2.4\times 10^{-5}$ a.u. is introduced corresponding to the interaction time of 1~ps (frequency resolution of 1~THz).

\begin{figure*}
  \begin{center}
  \includegraphics[width=0.45\textwidth]{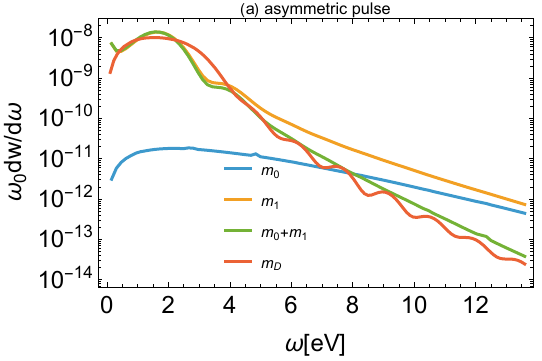}
\includegraphics[width=0.45\textwidth]{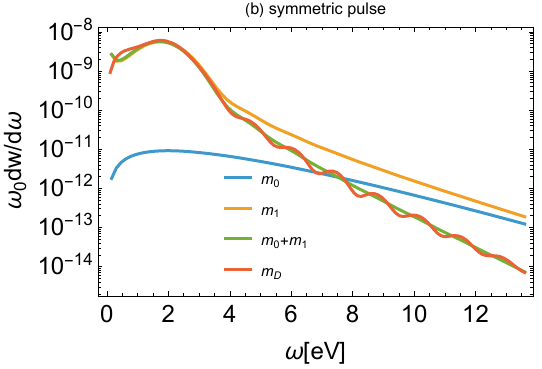}
\includegraphics[width=0.45\textwidth]{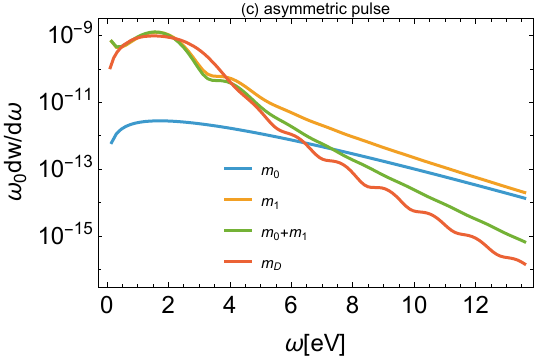}
\includegraphics[width=0.45\textwidth]{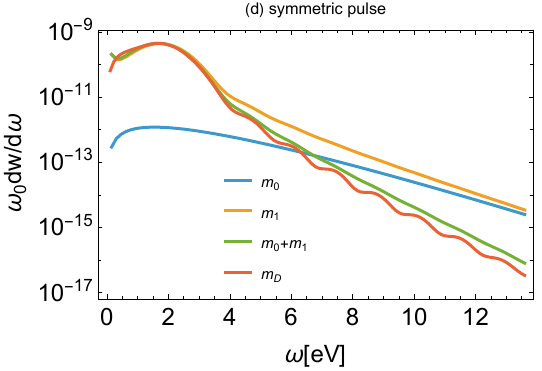}
   \caption{ Spontaneous radiation spectrum during tunneling ionization: (left column) asymmetric few-cycle laser pulse, (right column)  symmetric laser pulse, in both case $N=4$; (a,b) for $E_0=0.1$,  (c,d) for $E_0=0.075$. The contributions of amplitudes in different approximations $m_0$, $m_1$, and $m_0+m_1$ are specified. The results of the simple man Drude model are indicated by $m_D$.}
\label{MM}
\end{center}
\end{figure*}

\subsubsection{Radiation spectra}

The spectra in  symmetric and asymmetric few-cycle laser pulses with $E_0=0.1$, $\omega_0=0.05$, and the number of cycles $N=12$ in the 3D zero-range potential  are shown in Fig.~\ref{JJ}. In both cases the three-step HHG  dominates after the third harmonic, $\omega>3\omega_0\approx 4$ eV, while the continuum contribution via $J_{11}$ term does at low frequencies $\omega<4$ eV, see zoom of spectra for the case of the asymmetric laser pulse in Fig.~\ref{JJ2}. The first and second harmonics are mostly due to Thomson scattering, and the SFA calculation via  $J_{11}$  is well approximated by the Drude model in the interval $0.5<\omega<4$ eV. In particular, in the asymmetric pulse [Fig.~\ref{JJ2}], the  radiation has a spike in the near-zero-frequency  domain dominated by  the continuum  contribution ($J_{11}$), characteristic to the Drude model. Thus, we conclude that $J_{11}$ at low frequencies describes the Brunel radiation.

While Fig.~\ref{JJ} makes the impression that the $J_{11}$ induced radiation spectrum in the near-zero-frequency domain coincides with that via the Drude model (with the solely classical continuum contribution), the zoomed-in view of Fig.~\ref{JJ2} demonstrates   differences from the Drude model, especially at the peak of the near-zero-frequency region. 
The reason of the difference is that $J_{11}$ includes additionally the quantum free-free transition effect. The latter modifies the emission in most of the near-zero-frequency spectral region.  The comparison of the cases $\gamma=0.5$ and $\gamma=0.7$ in Fig.~\ref{JJ2}(c,d) shows that the quantum dynamics leads to small enhancement of the THz spectrum with respect to the classical Drude model in the nonadiabatic regime and vice versa  in the adiabatic regime.

The width of the near-zero-frequency spike is about $\Delta \omega  \sim 0.2$~eV, which is related to the laser pulse duration. In fact, the FWHM of the pulse in intensity is approximately $\tau\sim 200$ a.u. which corresponds to an energy of $1/\tau\sim 0.15$~eV,   roughly fitting to $\Delta \omega\sim 1/\tau$.

\section{Spontaneous radiation during tunneling ionization}\label{sec:spont}
\subsection{Spontaneous radiation spectra}

The spontaneous radiation during tunneling ionization is calculated as a photon emission during the transition between the state of the tunneling electron $\Psi(t)$ and the continuum state in the laser field:
\begin{eqnarray}
 dw =\left|m(\mathbf{k}, \mathbf{p})\right|^2d^3\mathbf{k}d^3\mathbf{p} ,
\end{eqnarray}
where the emission amplitude for the photon momentum $\mathbf{k}$, and the electron final momentum  $\mathbf{p}$ reads
\begin{eqnarray}
 m (\mathbf{k}, \mathbf{p})=\int dt \langle\psi_\mathbf{p}^{(f)}(t),1|\hat{\mathbf{p}}\cdot\hat{\mathbf{A}}(t)|\Psi (t),0\rangle,
\end{eqnarray}
with the SFA wave function of  the tunneling electron $\Psi(t)$. Integrating over the final electron momenta we obtain:
\begin{eqnarray}
\label{Wsp}
\frac{dw }{d\omega}&=&\frac{4\pi\omega^2 }{c^3}\int d^3\mathbf{p}\left|\int dt\langle\psi_\mathbf{p}^{(f)}(t),1|\hat{\mathbf{p}}\cdot\hat{\mathbf{A}}(t)|\Psi (t),0\rangle\right|^2.
\end{eqnarray}
In zeroth-order SFA $\Psi=\psi^{(0)}(t)$ and we obtain the photon emission probability which coincides with the result of Ref.~\cite{Milosevic_2023} in velocity gauge.
In the first-order SFA calculation, the electron initial state during the photon emission process is $\Psi(t)=\psi^{(0)}(t)+\psi^{(1)}(t)$, which includes the continuum dynamics in the tunneling ionization process via $\psi^{(1)}(t)$. Comparing the zeroth-order probability with the first-order one, we learn about the role of the continuum dynamics for the spontaneous radiation. We define the zero-order amplitude as
\begin{eqnarray}
 m_0(\mathbf{k}, \mathbf{p})=\int dt \langle\psi_\mathbf{p}^f(t),1|\hat{\mathbf{p}}\cdot\hat{\mathbf{A}}(t)|\psi^{(0)}(t),0\rangle,\\
\end{eqnarray}
and the first-order one
\begin{eqnarray}
 m_1(\mathbf{k}, \mathbf{p})=\int dt \langle\psi_\mathbf{p}^f(t),1|\hat{\mathbf{p}}\cdot\hat{\mathbf{A}}(t)|\psi^{(1)}(t),0\rangle,
\end{eqnarray}
as well as the total amplitude
\begin{eqnarray}
 m(\mathbf{k}, \mathbf{p})= m_0(\mathbf{k}, \mathbf{p})+ m_1(\mathbf{k}, \mathbf{p}).
\end{eqnarray}
The amplitudes are expressed in the following form:
\begin{eqnarray}
m_0(\mathbf{k},\mathbf{p})&=&-i\int dt\hat{\mathbf{A}}(t)\cdot\mathbf{m}_p^*(\mathbf{p},t)\\
 m_1(\mathbf{k},\mathbf{p})&=&\int dt'\left[\mathbf{p}+\mathbf{A}(t')\right]\cdot\hat{\mathbf{A}}(t')\int^{t'}dt m_x(\mathbf{p},t).\nonumber
\end{eqnarray}

The amplitude $m_1$ has a structure of  a nested integral and is transformed via two partial integrations to the following expression
\begin{eqnarray}
m_1&=&-\int dt'\left[\mathbf{p}+\mathbf{A}(t')\right]\cdot\hat{\mathbf{A}}(t')\,m_x(\mathbf{p},t')\int^{t'} dt\exp(i\omega t)\nonumber\\
&&-\int dt'm_x(\mathbf{p},t'')\int^{t''} dt' \mathbf{E}(t')\int^{t'} dt\exp(i\omega t).
\end{eqnarray}
The inner time integrals are analytical and the outer integration is performed numerically.
 The final momentum integration is also performed numerically.

The boundaries of the finite time integration introduce not physically relevant time diffraction oscillations in the integration result. To avoid it, we shift the time integration contour in the complex plane passing over the saddle point $t_s=i\gamma/\omega$, and omit the contribution of the vertical contours, which are responsible for the time diffraction, see a similar procedure in Ref.~ \cite{Klaiber_2022edge}. Thus, the boundary terms of the time integrals are treated like
\begin{eqnarray}
\int dt f(t)=\int ^{t_f+i\gamma/\omega}_{t_i+i \gamma/\omega}dt f(t).
\end{eqnarray}

We have carried out the calculation of the incoherent radiation spectra in the case of a symmetric and asymmetric  few-cycle laser pulses defined above with $E_0=0.1$ ($\gamma=0.5$) and $E_0=0.075$ ($\gamma=0.7$), $\omega_0=0.05$, and $N=4$. The spectra for the spontaneous radiation are shown in Fig.~\ref{MM}. The spectrum via $m_0(\mathbf{k}, \mathbf{p})$ is similar to the calculation of Ref.~\cite{Milosevic_2023}, with the spectral width $\Delta \omega \sim 1/\tau_K$, determined by the Keldysh time $\tau_K=\kappa/E_0$. Up to $\omega<6$~eV it is much smaller with respect to  $m_1(\mathbf{k}, \mathbf{p})$, which includes the continuum motion. In the near-zero-frequency region, $m_1(\mathbf{k}, \mathbf{p})$ has a spike similar to that in the coherent radiation. At larger frequencies $\omega>6$~eV,  $m_0(\mathbf{k}, \mathbf{p})$ and $m_1(\mathbf{k}, \mathbf{p})$  compensate each other. Thus, the amplitude $m_0(\mathbf{k}, \mathbf{p})$ does not describe the observable spectrum neither at low frequencies $\omega \rightarrow 0$, nor at high ones $\omega \rightarrow \infty$, but only near the cutoff $\omega\sim 8$ eV. Our remarkable observation is that there is no significant difference in spontaneous radiation in symmetric or asymmetric pulses. The qualitative behavior of the spectra does not change with increasing the Keldysh parameter, but the intensity of radiation decreases with larger $\gamma$ as the driving electric field decreases. The spectral width of the radiation is related to the Keldysh time $\Delta \omega \sim 1/\tau_K=\gamma/\omega$.

\subsection{Simple man Drude model for spontaneous radiation}

For an intuitive proof of the spontaneous radiation spectra, we have generalized the simple man Drude model of Eq.~(\ref{Bab}) for the spontaneous radiation described by Eq.~(\ref{Wsp}). Taking into account that in the strong-field ionization three-step model~\cite{Corkum_1993} there is a mapping between the ionization time and the momentum,
and substituting $ \langle\psi_\mathbf{p}^{(f)}(t) |\hat{\mathbf{p}} |\Psi (t) \rangle \rightarrow \mathbf{v}(t,t')$, as well as
$ \langle 1| \hat{\mathbf{A}}(t)| 0\rangle \rightarrow \langle\hat{\mathbf{A}}(t)\rangle \equiv -1/(2\pi\sqrt{\omega})\hat{\mathbf{e}}e^{i\omega t}$,
we arrive at the simpleman Drude model representation for  the spontaneous radiation of Eq.~(\ref{Wsp}):
\begin{eqnarray}
\frac{dw_{\rm{D}}}{d\omega}=\frac{4\pi\omega^2}{ c^3}\int dt' \left|\int^{t_f+i\gamma/\omega}_{t'+i\gamma/\omega} dt \mathbf{v}(t,t')\cdot\langle\hat{\mathbf{A}}(t)\rangle\right|^2W(t').\nonumber\\
\end{eqnarray}
The results via the Drude model (via the current $J_D$) for the spontaneous radiation spectra are also presented in Fig.~\ref{MM}. The Drude model is qualitatively in accordance with $m_{1}$ in the low frequency region and confirms the enhancement of spectra in this frequency region  in contrast to the behavior of the spectra based on the direct ionization $m_0$ amplitude. At large frequencies $\omega>6$ eV, when the radiation is dominated by interference of $m_0$ and $m_1$, the Drude model still surprisingly agrees quite well with the SFA-calculation.

\subsection{Correlated electron and photon measurement}
\begin{figure}[b]
  \begin{center}
\includegraphics[width=0.5\textwidth]{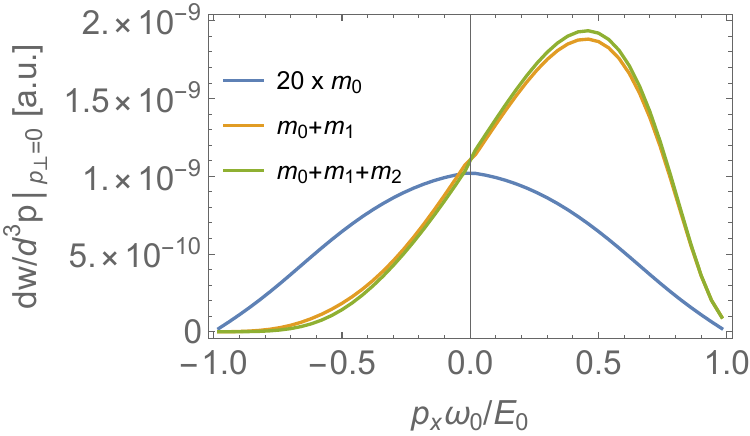}
 \caption{Total probability of photon emission vs electron final momenta in the case of  the correlated electron and photon measurement, for $E_0=0.2$ a.u. and a half-cycle pulse; the value for the $m_0$ contribution is multiplied by $20$. }
  \label{tt2}
\end{center}
\end{figure}

 The under-the-barrier recollisions are described by the $m_2$ amplitude. Our calculations show that its contribution to the coherent and spontaneous radiation spectral distribution is negligible, i.e., the radiation spectra do not carry information about under-the-barrier recollisions. Is there any other way to see this signature in the radiation properties? To answer this question, we discuss the correlated electron and photon measurement.

The spontaneous emission spectra in Fig.~\ref{MM} have been calculated integrating over the final electron momentum. However, when  the correlated electron and photon measurement is carried out in an experiment, one can access the electron momentum resolved photon spectra.  The electron momentum resolved total photon yield in a half-cycle laser pulse in the case of the correlated measurement is defined as follows:
\begin{eqnarray}
\frac{dw}{d^3\mathbf{p}}=\frac{4\pi}{c^3}\int \omega^2 d\omega \left| m_0(\mathbf{k}, \mathbf{p})+ m_1(\mathbf{k}, \mathbf{p})  \right|^2.\label{correlated_yield}
\end{eqnarray}
The zeroth-order amplitude
\begin{eqnarray}
m_0(\mathbf{k},\mathbf{p})=\int dt \hat{\mathbf{A}}(t)\cdot\mathbf{m}_p(\mathbf{p},t),
\end{eqnarray}
is calculated via  the saddle-point integration in time $t$, where the saddle points are found numerically in a half-cycle sinusoidal laser pulse.
For the calculation of the first-order amplitude,  the Drude model for spontaneous emission is applied:
\begin{eqnarray}
\label{m1_corr}
m_1(\mathbf{k},p_x,p_\perp =0 )=\int^{t_f+i\gamma/\omega}_{t'+i\gamma/\omega} dt \mathbf{v}(t,t')\cdot\langle\hat{\mathbf{A}}(t)\rangle\sqrt{W(t',p_\perp=0)}\nonumber\\
\end{eqnarray}
with the time-to-momentum mapping $p_x=-A(t')$, and $W(t,p_\perp=0)=W(t)\kappa/(\pi|\mathbf{E}(t)|)$. The $t$-integration in $m_1$, and $\omega$-integration in Eq.~(\ref{correlated_yield}) are carried out numerically.

We include the second-order correction to the emission amplitude $m_2$ heuristically,  taking into account the  correction in the ionization rate, where we replace the binding energy of the initial state $-\kappa^2/2$ by $-(\kappa^2/2)\{1+i\exp[-(2/3)(\kappa^3/|\mathbf{E}(t)|)]\}$ \cite{Klaiber_2024}.

The correlated photon yield is presented in Fig.~\ref{tt2}. The higher order corrections to the spectrum via $m_0$ and $m_1$ induce a shift of the position of the maximum. This is a result of the continuum motion. While in the given half-cycle laser pulse the photoelectron momentum peak is shifted from the zero momentum, the momentum shift $\delta p$ in the photoelectron momentum resolved photon yield  is smaller, because it is additionally multiplied in Eq.~(\ref{m1_corr}) by the Fourier component of the ionizing electron velocity. The  momentum shift  approximately scales as $\delta p\sim (E_0/\kappa^3)^{2/3}(E_0/\omega_0)$ according to the fit to the numerical calculations.

The small contribution of the under-the-barrier recollisions via $m_2$ can be distinguished near the peak. However, this does not lead to any qualitative change of the spectrum.

\begin{figure} [b]
  \begin{center}
\includegraphics[width=0.45\textwidth]{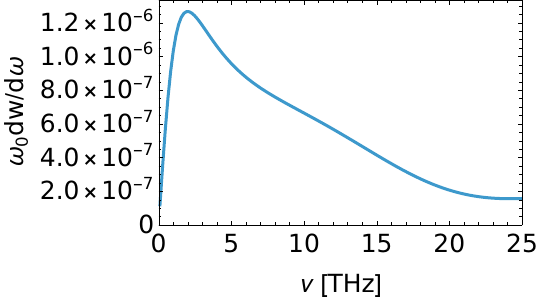}
 \caption{The radiation spectrum at strong field ionization  near the low-frequency region for the  approximate parameters of the experiment of Ref.~\cite{Clerici_2013}: $E_0=0.1$, ${\cal A}=0.2$.}
  \label{kim}
\end{center}
\end{figure}

\section{Comparison with experimental results}\label{sec:experiment}

The main distinguishing feature of the Brunel radiation is a large spike at near zero frequencies. To validate out theory based on SFA, we compare its result with an example of the experimental spectrum of Ref.~\cite{Clerici_2013}. The laser parameters employed in this experiment are: the wavelength $\lambda=800$ nm, pulse duration $\tau=60$ fs, pulse energy $\varepsilon=400\,\mu$J. Then,  the average intensity can be estimated as $I_0 \approx 3.3\times 10^{14} \rm W/cm^2$, or the laser field is $E_0\approx 0.1$ a.u.
The second harmonic energy is approximately 5\%, then  the asymmetry ${\cal A}=\sqrt{0.05}\approx 0.2 $. The radiation spectrum in the nitrogen gas with the laser parameters approximately deduced from the provided information is shown in Fig.~\ref{kim} in the low-frequency region. We see  an  order of magnitude decrease of the radiation intensity in the spectral region up to 20 THz. A similar spike   existed in the experimental spectrum of Fig.~3 in Ref.~\cite{Clerici_2013}. Thus, our simple SFA theory qualitatively reproduces the main feature of the experimental spectrum. There exist deviations in details from the experimental result, for which more accurate description is necessary, in particular the inclusion of explicit atomic potential, rather than the short-range one as in our SFA theory.

\section{Estimation of radiation intensities}\label{sec:estim}

Finally, let us give an order of magnitude estimations of the coherent and incoherent Brunel radiation.
We proceed with the dipole radiation power~\cite{Landau_2}:
\begin{eqnarray}
\label{dipole}
 I = \frac{2 \ddot{d }^2}{ 3c^3} .
\end{eqnarray}
Estimating the current from Eq.~(\ref{Bab}):
\begin{eqnarray}
\dot{d}_{\rm coh}\sim (E_0/\omega_0) w_i T,
\end{eqnarray}
where $w_i\sim   (\kappa^2/2)  (E_0/E_a)\exp[-2E_a/(3E_0)]$ is the laser  ionization rate, and $T=2\pi /\omega_0$ the laser period. The probability of the coherent Brunel photon emission per laser period with the energy $\omega$  is
 \begin{eqnarray}
 w_{\rm coh}\sim \frac{I_{\rm coh} T}{\omega } \sim \frac{ \alpha E_0^2}{c^2\omega}\left(\frac{\omega}{\omega_0}\right)^2w_i^2T^3\sim \alpha a_0^2\gamma^{-1}(w_iT)^2,
\end{eqnarray}
where $a_0=E_0/c\omega_0$ is   the relativistic invariant field  parameter.

Now let us estimate the probability of the spontaneous Brunel radiation again via the dipole radiation theory. For spontaneous radiation, we estimate the dipole moment derivative as
\begin{eqnarray}
\dot{d}_{\rm sp}\sim \omega_c \ell_c ,
\end{eqnarray}
with the characteristic frequency $\omega_c\sim 1/\tau_K$, and $\ell_c \sim \kappa \tau_K$, with the Keldysh time $\tau_K=\kappa/E_0$, and the atomic velocity $\kappa$. Then, the spontaneous radiation power is
\begin{eqnarray}
 I_{\rm sp}\sim  \frac{\alpha \omega_{c}^4 \ell^2}{c^2}w_i T,
\end{eqnarray}
where we multiplied by the ionization probability because the dipole moment $d_{\rm sp}$ is created during tunneling. Then, the probability of the spontaneous photon emission during tunneling ionization:
 \begin{eqnarray}
 w_{\rm sp}\sim  \frac{I_{\rm sp} T}{\omega_c}w_i T\sim \frac{\alpha \omega_{c}^3 \ell^2}{c^2}w_i T^2\sim \alpha \frac{\kappa^2}{c^2} \gamma^{-1} w_i T,
\end{eqnarray}
with the Keldysh parameter $\gamma=\omega_0\tau_K$.

Thus, the ratio of the probabilities is
 \begin{eqnarray}
 \frac{w_{\rm sp}}{ w_{\rm coh}}\sim  \frac{\kappa^2}{c^2} \frac{1}{a_0^2 w_i T}
\end{eqnarray}
For $E_0=0.1$ (the laser intensity $3.5\times 10^{14}$ W/cm$^2$),  $E_0/E_a\sim 0.1$, and $a_0\sim 1.5\times 10^{-2}$,  $\frac{I_p}{\omega_0}\sim 10$, $e^{-\frac{2E_a}{3E_0}}\sim  1.3\times 10^{-3}$, $w_i T\sim 10^{-2}$, we obtain $ w_{\rm sp}/ w_{\rm coh}\sim 10^2$.  The latter ratio is  roughly in accordance with spectra in Figs.~\ref{JJ} and \ref{MM}.

We may estimate also the spontaneous radiation per electron: $w_{\rm sp}^{(e)}\sim w_{\rm sp}/w_i T\sim (\alpha/\gamma) (\kappa/c)^2\sim 10^{-6}$, which indicates complications for the electron and photon coincidence measurement. The Brunel radiation scales as $1/\gamma$ with the Keldysh parameter.

When the Brunel radiation is observed with a gas target,  the coherent Brunel radiation will be dominating over the spontaneous one, if the target density is large enough: ${\cal N}=n{\cal V}>  w_{\rm sp}/ w_{\rm coh} $, with the particle number density $n$, the focal volume ${\cal V}\sim (\pi w^2/\lambda)(\pi w^2)$, and the laser beam waist size $w$. With $w\sim 3\lambda$, $\lambda\sim 1\,\mu$, this will take place at a gas density $n>10^{11}$ cm$^{-3}$.

\section{Conclusion}\label{sec:concl}

We have investigated the coherent and incoherent radiation during tunneling ionization. The coherent radiation in the near-zero-frequency THz  domain ($\omega \sim 0.2$ eV) is dominated by the continuum dynamics via the quantum free-free transition $J_{11}$ term in the current. We have shown that due to the quantum dynamics the radiation is modified especially in the adiabatic limit.

The spontaneous radiation is not affected by the asymmetry of the driving laser pulse. The characteristic width of the spontaneous radiation is determined by the Keldysh time $\Delta \omega \sim 1/\tau_K$. There are two contributions into the spontaneous radiation: due to the direct transition during tunneling (with the amplitude $m_0$) investigated in Ref~\cite{Milosevic_2023}, and due to additional motion in the continuum (with  the amplitude $m_1$). At larger frequencies $\omega \gtrsim 1/\tau_K$, $m_0$ and $m_1$ terms compensate each other and the total radiation is suppressed. At low frequencies $\omega \lesssim 1/\tau_K$, the radiation is dominated by the continuum dynamics via the amplitude $m_1$.  Thus, the contribution of the direct transition in the radiation essentially  describes the spontaneous radiation only near the cutoff.

\bibliography{strong_fields_bibliography}

 \end{document}